\documentclass[11pt]{article}
 \usepackage{amsfonts}
 \usepackage{amssymb}
 \def\bsh{\backslash}
 \def\bdt{\dot \beta}
 \def\adt{\dot \alpha}

 \newfont{\bbbold}{msbm10 scaled \magstep1}

 \def\bbC{\mbox{\bbbold C}}

 \def\bbF{\mbox{\bbbold F}}
 \def\bbG{\mbox{\bbbold G}}

 \def\bbR{\mbox{\bbbold R}}

 \def\cN{{\cal N}}

 \def\cV{{\cal V}}

 \newfont{\goth}{eufm10 scaled \magstep1}
 
 \def\gb{\mbox{\goth b}}

 \def\gg{\mbox{\goth g}}

 \def\gl{\mbox{\goth l}}

 \def\gp{\mbox{\goth p}}

 \def\gs{\mbox{\goth s}}
 
 \def\gu{\mbox{\goth u}}

 \def\a{\alpha}
 \def\b{\beta}
 \def\c{\gamma}\def\cdt{\dot\gamma}
 \def\d{\delta}\def\ddt{\dot\delta}
 \def\e{\epsilon}\def\vare{\varepsilon}
 \def\vf{\varphi}
 \def\z{\zeta}
 \def\k{\kappa}
 \def\l{\lambda}

 \def\th{\theta}
 
 \def\be{\begin{equation}}\def\ee{\end{equation}}
 \def\bea{\begin{eqnarray}}\def\eea{\end{eqnarray}}
 \def\ba{\begin{array}}\def\ea{\end{array}}

 \def\del{\partial}

 \def\xz{\times}

 \def\del{\partial}
 \def\bfk{{\bf k}}\def\bfK{{\bf K}}
 \def\bt{\bullet}

 \let\la=\label

 {}

 \def\nn{\nonumber}
 \def\bd{\begin{document}}
 \def\ed{\end{document}}
 \def\bea{\begin{eqnarray}}\def\barr{\begin{array}}\def\earr{\end{array}}
 \def\eea{\end{eqnarray}}
 \def\ft#1#2{{\textstyle{{\scriptstyle #1}\over {\scriptstyle #2}}}}
 \def\fft#1#2{{#1 \over #2}}
 \newcommand{\eq}[1]{(\ref{#1})}
 \def\eqs#1#2{(\ref{#1}-\ref{#2})}
 \def\det{{\rm det\,}}
 \def\tr{{\rm tr}}\def\Tr{{\rm Tr}}

\begin{document}

 \thispagestyle{empty}

 \hfill{KCL-TH-00-22}

 \hfill{hep-th/0005135}

 \hfill{\today}

 \vspace{20pt}

 \begin{center}
 {\Large{\bf On Harmonic Superspaces and Superconformal Fields in
 Four Dimensions}}
 \vspace{30pt}

 {P. Heslop and P.S. Howe} \vskip 1cm {Department of Mathematics}
 \vskip 1cm {King's College, London} \vspace{15pt}

 \vspace{60pt}

 {\bf Abstract}

 \end{center}

 Representations of four-dimensional superconformal groups on harmonic
 superfields are discussed. It is shown how various short
 representations can be obtained  by parabolic induction.
 It is also shown that such short multiplets may admit several descriptions as 
 superfields
 on different superspaces. In particular, this is the case
 for on-shell massless superfields. This allows a description of short
 representations as explicit products of fundamental fields. Superconformal
 transformations of analytic fields in real harmonic superspaces are
 given explicitly.
 {\vfill\leftline{}\vfill \vskip  10pt

 \baselineskip=15pt \pagebreak \setcounter{page}{1}

 \section{Introduction}

 The irreducible representations of the four-dimensional
 superconformal group $SU(2,2|N)$ have been much studied in the
 literature \cite{screp}. The theory of such representations has assumed
 some
 significance recently in the light of the Maldacena conjecture \cite{mal}
 relating IIB supergravity and string theory on backgrounds with an
 $AdS_5$ factor to four-dimensional superconformal field theories on
 the boundary. A particularly important class of operators that can
 arise consists of those operators which correspond to short representations of the
 superconformal group since these are expected to be protected from
 quantum corrections and thus not acquire anomalous dimensions. (See, for example, \cite{af1,af2,fz}).

 One way of constructing representations explicitly is the method of
 parabolic induction. In the context of superconformal groups this was
 discussed in some detail in \cite{hh1}, although a direct comparison
 with the more algebraic group-theoretic
 results of \cite{screp} was not made at the time. However, it was shown how certain
 ultra-short representations, which correspond to on-shell massless
 supermultiplets and which can be described by constrained superfields
 in ordinary $N$-extended Minkowski superspace \cite{s,hst}, can be
 described by
 ``generalised chiral'' or ``Grassmann analytic'' (G-analytic)
 superfields in
 harmonic superspaces \cite{hh2}. A simple example of a composite operator - the
 $N=4$ supercurrent - was also constructed as the trace of the square of the $N=4$
 Yang-Mills field-strength superfield. In \cite{hw} it was shown
 that
 there is in fact a
 series of such G-analytic superfields obtained by simply taking
 the trace of powers of the field strength, and it was conjectured that this circumstance might lead to restrictions on the correlations functions of such operators, an idea that has been exploited using both $N=4$ and $N=2$ superspaces in refs \cite{anal}.
These
 composites all have maximum spin 2 and it was subsequently realised
 in \cite{af1} that the superconformal representations which are
 realised this
 way are dual to the bulk Kaluza-Klein representations which arise in
 the compactification of IIB supergravity on $AdS_5\xz S^5$ \cite{gun}.
 In the case of an Abelian field strength this
 construction can easily be
 generalised to the case of the $D=6,
 (2,0)$ tensor multiplet and the $D=3, N=8$ scalar multiplet, these
 being the multiplets which are relevant to M-theory compactified on
 $AdS_4\xz S^7$ and $AdS_7\xz S^4$ respectively \cite{h}
 (for other discussions of $D=6, (2,0)$ representations see \cite{sei,roz,gun2}).

 In a recent series of papers short representations in $D=4$
 \cite{afsz, fs1},
 $D=6$ \cite{fs2} and $D=3$ \cite{fs3} have been discussed in the context of 
 harmonic superspaces which have the form of super Minkowski space times
 an internal
 coset space with isotropy group equal to the maximal
 torus in the internal symmetry group. As well as
 reproducing the KK
 multiplets the authors of these papers were able to obtain many other short
 representations
 by taking products of the underlying massless multiplets and then
 imposing harmonic analyticity on the outcome.

 In the current paper we adopt a more general approach which
 makes use of all of the harmonic superspaces which are compatible with
 superconformal symmetry. We first discuss the short representations directly
 using the method of parabolic induction. To do this it is necessary to
 complexify spacetime and the superconformal group since the
 superspaces we are interested in are not cosets of the real
 superconformal group. One advantage of this approach is that
 superconformal symmetry is manifest throughout.

The short representations act on superfields which depend on fewer
odd coordinates than are present in $N$-extended super Minkowski
space; such G-analytic superfields obey constraints which
generalise the constraints satisfied by chiral superfields. We show
that each short representation generically admits 
realisations on different homogeneous superspaces which incorporate
different numbers of G-analyticity constraints. Amongst these
realisations there are those which are ``maximally efficient'' in
that the maximum number of G-analyticity constraints compatible
with the representation is explicitly incorporated. It will be seen
that the number of G-analyticity constraints of a given realisation
places constraints on the geometry of the internal manifold of the
associated harmonic superspace. The fewer G-analyticity constraints
there are (equivalently, the larger the number of odd coordinates
of the superspace concerned) the larger the number of
representations that will be compatible with it. Thus,
G-analyticity of type (1,1) is compatible with all representations
of series C while G-analyticity of type (1,0) (or (0,1)) is
compatible with all short representations (see, for example,
\cite{fz,afsz} for discussions of these representations).

After a brief review of homogeneous superspaces of the
superconformal group and parabolic induction, we discuss the short
representations in detail in section 3. In section 4, we discuss
on-shell massless supermultiplets from this viewpoint; these are
special superconformal fields in that they carry ultra-short
representations. Each of these massless supermultiplets can be
realised in many different ways in general and this allows us to
construct general representations out of them by taking products on
appropriate superspaces. This approach has the virtue of not
requiring the projection step of \cite{afsz,fs1} - the product
superfields are naturally superconformal fields on the appropriate
superspace. In particular, in non-Abelian $N=4$ supersymmetric
Yang-Mills theory the single, double and triple trace
gauge-invariant composite operators can be constructed
straightforwardly as G-analytic and harmonic analytic superfields
by making use of the three different realisations of the underlying
field strength multiplet. Aspects of multiple trace operators, which are believed to correspond to multiparticle states in the AdS/CFT context, have been studied in \cite{mult} and in harmonic superspace in \cite{afsz}.

The paper concludes in section 5 with an explicit discussion of
superconformal transformations acting on shortened supermultiplets
defined as analytic superfields on real harmonic superspaces.

 \section{Preliminaries}

 \subsection{Harmonic superspace}

 Harmonic superspaces are usually taken to be superspaces of the form
 $M\xz F$ where $M$ is super Minkowski space and $F=H\bsh G$ is a coset
 space of the internal symmetry
 group $G$ with isotropy group $H$ \cite{gikos1,gikos2,gikos3,hh1,hh2}.
 Usually $F$ is a compact, complex
 manifold; in $N$-extended supersymmetry in four dimensions it can be
 taken to be of the form $\bbF_{\bfk}(N)=S\left(U(k_1)\xz
 U(k_2-k_1) \ldots U(N-k_{\ell})\right)\bsh SU(N)$ where $\bfk:=(k_1,\ldots
 k_{\ell})$ is a sequence of positive integers with $k_1<k_2\ldots
 k_{\ell}<N$ \cite{ikno,hh1,hh2}. The space $\bbF_{\bfk}(N)$ is a
 flag manifold: a point in
 $\bbF_{\bfk}(N)$ is a flag in $\bbC^N$, i.e. a sequence of subspaces
 $\{V_{i}\}$, $\dim\,V_i=k_i$, such that $V_{1}\subset V_{2} \ldots
 V_{{\ell}}\subset \bbC^N$. Special cases of flag manifolds include
 projective spaces and Grassmannians. A complete flag is one with
 $\bfk=(1,2,3,\ldots N-1)$. The space of complete flags, which we
 shall denote $\bbF_B$, is $U(1)^{N-1}\bsh SU(N)$; it can also be
 thought of as the coset space $B\bsh SL(N,\bbC)$ where $B$ is the
 Borel subgroup of $SL(N)$.

 In the GIKOS formalism \cite{gikos1} one works with fields defined
 on $M\xz G$
 which are equivariant with respect to $H$. That is, one has fields
 $A(z,u),\ z\in M, u\in G$ which take their values in some
 representation space of $H$ and which satisfy $A(z,hu)= R(h)A(z,u)$,
 $R$ being the representation of $H$ in question. Such fields are
 equivalent to sections of vector bundles over harmonic superspace. In
 index notation we write $u_I{}^i$ where the upper case index
 indicates a representation of $H$ and the lower case index the
 defining representation of $G$, and where both indices run from $1$ to $N$.
 The coordinates of Minkowski
 superspace are denoted by $(x^{\a\adt},\th^{\a i},\bar\th^{\adt}_i)$
 where $\a$ and $\adt$ are two-component spinor indices. The covariant
 derivatives $(\del_{\a\adt},D_{\a i},\bar D_{\adt}^i)$ satisfy

 \be
 [D_{\a i},\bar D_{\adt}^j]=i\d_i{}^j\del_{\a\adt} \ee

 with all other (anti)-commutators being zero. With the aid of $u$ we
 can convert $G$ indices to $H$ indices; this means that we can single
 out subsets of $D$'s and $\bar D$'s which anticommute with each other
 and which can therefore consistently be taken to annihilate
 superfields. Let us suppose that the index $I$ splits as follows
 under $H$: $I=(r,r'',r')$, where
 $r=1,\ldots p,\ r''=p+1,\ldots N-q,\ r'=N-q+1,\ldots N$,
 and where $(p+q)\leq N$, then we can have superfields $A$ which obey
 the constraints

 \be
 D_{\a r}A=\bar D_{\adt}^{r'}A=0
\la{pq} \ee

 where

 \be D_{\a r}:=u_r{}^i D_{\a i};\qquad \bar D_{\adt}^{r'}=\bar
 D_{\adt}^i(u^{-1})_i{}^{r'} \ee

 Constraints such as these are called Grassmann analyticity
 (G-analyticity) conditions following \cite{gio}; they generalise the notion
 of chirality which was first introduced in the context of $N=1$
 supersymmetry in \cite{fwz}. We shall refer to the G-analyticity conditions
 of equation \eq{pq} as
 being of type $(p,q)$. The space of G-analytic structures of type
 $(p,q)$ on $N$-extended Minkowski superspace (in four dimensions) is
 the flag manifold
 $\bbF_{p,N-q}(N)=S(U(p)\xz U(N-(p+q))\xz U(q))\bsh SU(N)$,
 and
 the harmonic superspace with this internal coset is referred to as
 $(N,p,q)$ harmonic superspace. Later on we shall make a refinement of
 this definition. The derivatives on $G$ can be taken to be the
 right-invariant vector fields, $D_I{}^J,\ D_I{}^I=0$. They act as
 follows:

 \be
 D_I{}^J u_K{}^k=\d_K{}^J u_I{}^k-{1\over N}\d_I{}^J u_K{}^k \ee

 and they obey the reality condition

 \be
 \bar D^J{}_I=-D_I{}^J \ee

 These derivatives divide into three sets: the set
 $\{D_{r}{}^{s'},D_{r}{}^{s''},D_{r''}{}^{s'}\}$, corresponding to the
 $\bar\del$ operator on the complex coset space (i.e. holomorphic
 functions are annihilated by $\bar\del$), the complex conjugate set
 $\{D_{r'}{}^{s},D_{r''}{}^{s},D_{r'}{}^{s''}\}$ which corresponds to
 the $\del$ operator, and the set
 $\{D_{r}{}^{s},D_{r'}{}^{s'},D_{r''}{}^{s''}\}$ which corresponds to
 the isotropy algebra. As well as imposing G-analyticity we can impose
 harmonic analyticity (H-analyticity) by demanding that a superfield
 be annihilated by the first of these sets of derivatives. Indeed,
 since the algebra of all of these operators and the spinorial
 derivatives defining G-analyticity closes, it is consistent to impose
 both G-analyticity and H-analyticity on the same superfield. Such
 superfields will be referred to as CR-analytic, or simply analytic.
 It was observed in \cite{rs} that analyticity conditions such as these
 define a generalisation of a complex structure, known as a CR
 structure, on harmonic superspace. In this language, a G-analytic
 structure can be
 considered to be a CR-structure which is purely odd.

 \subsection{Harmonic conjugation}

 It was noted in the first paper on harmonic superspace
 \cite{gikos1} that it is necessary to use a modified conjugation
 in order to handle real representations of $SU(2)$ in a
 straightforward manner. This conjugation combines the
 antipodal map on the internal manifold $\bbC P^1$ with ordinary
 complex conjugation. This was generalised to $(N,p,q)=(3,1,1)$ in
 \cite{gikos2} and to any $(N,p,p)$ space in \cite{hh2}. In a
 general $(N,p,q)$ superspace, with internal manifold
 $\bbF_{p,N-q}(N)$ the analogue of the antipodal map becomes a map
 from $\bbF_{p,N-q}(N)$ to $\bbF_{q,N-p}(N)$; if we combine this
 transformation with complex conjugation we obtain a map which maps
 the CR-structure on $(N,p,q)$ harmonic superspace to the CR
 structure on $(N,q,p)$ superspace. We shall refer to this as
 harmonic conjugation.

 Working on $M\xz SU(N)$ we write $u_I{}^i$ as before as
 $(u_r{}^i,u_{r''}{}^i,u_{r'}{}^i)$ with $r=1,\ldots p$ and
 $r'=N-q+1,\ldots N$. This is the appropriate splitting of $I$ for
 $(N,p,q)$ superspace. We denote the splitting relevant to $(N,q,p)$ by
 $I=(R,R'',R')$ with $R=1,\ldots q,R'=N-p+1,\ldots N$. The mapping we start with is given by

 \be
 u\mapsto k u
 \ee

 where $k$ is the matrix

 \be
 k=\left(\ba{ccc}
 0&0&-1_q\\
 0&1_r&0\\
 1_p&0&0
 \ea\right)
 \ee

 It is easy to check that $k h k^{-1}\in S(U(q)\xz U(N-(p+q))\xz
 U(p))$ provided that $h\in S(U(p)\xz U(N-(p+q))\xz U(q))$, so that this does
 indeed define a map from one coset space to the other, i.e. from
 $\bbF_{p,N-q}(N)$ to $\bbF_{q,N-p}(N)$. If we combine this with
 complex conjugation we find

 \bea
 u_{r}{}^i &\mapsto& (u^{-1})_i{}^{R'} \nn\\
 u_{r''}{}^i &\mapsto& (u^{-1})_i{}^{R''}\nn\\
 u_{r'}{}^i &\mapsto& -(u^{-1})_i{}^{R}
 \eea

 For the inverses we have

\bea
 (u^{-1})_{i}{}^{r} &\mapsto& u_{R'}{}^i \nn\\
 (u^{-1})_{i}{}^{r''} &\mapsto& u_{R''}{}^{i}\nn\\
 (u^{-1})_{i}{}^{r'} &\mapsto& -u_{R}{}^{i}
 \eea

We then find

\bea D_{\a r}&\mapsto& \bar D_{\adt}^{R'}\nn\\ \bar
D_{\adt}^{r'}&\mapsto& -D_{\a R} \eea

while

\bea D_{r}{}^{s'}&\mapsto& D_{S}{}^{R'}\nn\\
D_{r}{}^{s''}&\mapsto&
-D_{S''}{}^{R'}\nn\\
D_{r''}{}^{s'}&\mapsto& D_{S}{}^{R''}
\eea

so that the derivatives of the $(N,p,q)$ CR-structure do indeed
transform into those of the $(N,q,p)$ CR-structure, and therefore
analytic fields on $(N,p,q)$ superspace transform into analytic
fields on $(N,q,p)$ superspace.

 \subsection{Complex superspaces}

 It is the analytic superfields defined on harmonic superspaces of the
 above type, or related ones, that we shall be interested in for the
 description of short representations of the superconformal group.
 However, from a group-theoretical point of view, it is convenient
 to take a slightly
 different approach to harmonic superspaces, namely as coset spaces of
 the superconformal group. A slight technical problem presents itself
 here, because the spaces we are ultimately interested in, i.e. the
 so-called analytic superspaces where the G-analyticity constraints
 are automatically solved, are not coset spaces of the real
 superconformal group
 $SU(2,2|N)$; we can circumvent this problem by complexifying both the group and
 spacetime. The fields we are interested in will then be holomorphic
 fields defined on coset spaces of the complexified superconformal
 group $SL(4|N)$ (we omit the $\bbC$ which is always understood). The
 isotropy groups will be parabolic subgroups, which means that the
 superspaces themselves will be flag supermanifolds, that is, spaces
 whose points correspond to nested sequences of sub-supervector spaces
 of $\bbC^{4|N}$. The internal parts of these spaces will, however, be
 the same as in the real case above, the only difference being that we
 treat them exclusively as complex manifolds. When we restrict spacetime
 and the
 odd coordinates to be real we recover analytic fields of the
 type we have discussed above on $(N,p,q)$ harmonic superspaces. The
 virtue of this approach is that superconformal
 covariance is manifest.

 We begin by considering complexified Minkowski space which can be
 viewed as an open subset of the coset space $P\bsh SL(4)$, $SL(4)$
 being the complexified conformal group and $P$ the subgroup of
 matrices of the following shape:

 \be
 \left(\ba{cccc} \bullet&\bullet&&\\ \bullet&\bullet&&\\
 \bullet&\bullet&\bullet&\bullet\\ \bullet&\bullet&\bullet&\bullet \ea
 \right) \ee

 where the bullets denote elements which do not have to be zero. The
 blank region can be thought of as corresponding to spacetime.  Indeed,
 we can choose a coset representative of the form

 \be
 M\ni x\mapsto s(x)= \left(\ba{cc} 1_2&x\\ 0_2&1_2 \ea \right) \ee

 where each entry is a two-by-two matrix. From this one can easily
 work out that the transformation of $x$ under the conformal group is
 the usual one by using standard homogeneous space techniques.

 Complexified super Minkowski space has the form $P\bsh SL(4|N)$ where
 $P$ consists of matrices of the form

 \be
 \left( \ba{cccc|ccc}
  \bullet & \bullet &&&&&\\
  \bullet & \bullet &&&&&\\
  \bullet & \bullet &\bullet &\bullet &\bullet &. &\bullet\\
  \bullet & \bullet &\bullet &\bullet &\bullet &. &\bullet\\
 \hline
  \bullet &\bullet &&& \bullet &. &\bullet\\
  \bullet &\bullet &&& . &. &.\\
  \bullet &\bullet &&& \bullet &. &\bullet
 \ea \right)
 \ee

 The standard coset representative is

 \be
  M\ni z\mapsto s(z)=\left( \ba{cc|ccc}
  1_2&x &&\th&  \\
  0&1_2 &&0&\\
  \hline
  &&&&\\
  0&\vf &&1_N&\\
  &&&&
  \ea\right)
 \ee

 where $\vf$ denotes the $N$ dotted spinorial coordinates
 which become
 the complex conjugates of the $\th$'s in the real case.

 Complexified $(N,p,q)$ harmonic superspace is given by the
 following subgroup

 \be
 \begin{picture}(300,200)(20,-100)
 \put(10,0){$ \left( \ba{cccc|ccccccccc}
 \bt&\bt&&& &&&&&&&&\\
 \bt&\bt&&& &&&&&&&&\\
 \bt&\bt&\bt&\bt&  \bt&.&\bt&\bt&.&\bt&\bt&.&\bt\\
 \bt&\bt&\bt&\bt&  \bt&.&\bt&\bt&.&\bt&\bt&.&\bt\\
   \hline
 \bt&\bt&&&  \bt&.&\bt &&&&&&\\ .&.&&& .&.&.&&&&&&\\ \bt&\bt&&&
 \bt&.&\bt &&&&&&\\ \bt&\bt&&& \bt&.&\bt&\bt&.&\bt &&&\\ .&.&&&
 .&.&.&.&.&. &&&\\ \bt&\bt&&& \bt&.&\bt&\bt&.&\bt &&&\\
 \bt&\bt&&&  \bt&.&\bt&\bt&.&\bt&\bt&.&\bt\\ \bt&\bt&&&
 \bt&.&\bt&\bt&.&\bt&\bt&.&\bt\\ \bt&\bt&\bt&\bt&
 \bt&.&\bt&\bt&.&\bt&\bt&.&\bt
 \end{array} \right)$}
  \put(110,10) {$
 \left. \phantom{\begin{array}{c} \bt\\ .
 \\ \bt \end{array} } \right\}{\scriptscriptstyle p} $}
 \put(210,-70) {$ \left. \phantom{\begin{array}{c}  \\
 \\ \end{array} } \right\}{\scriptscriptstyle q} $} \put(155,-30) {$
 \left. \phantom{\begin{array}{c} \bt \\ . \\ \bt
 \end{array} } \right\}{\scriptscriptstyle N-p-q} $}
 \end{picture}
 \ee

Locally, this space has the form of complex super Minkowski space
times the internal flag manifold $\bbF_{p,N-q}(N)$. The related
$(N,p,q)$ analytic superspace has the same body but fewer odd
coordinates; the relevant subgroup is

 \be
 \begin{picture}(300,200)(20,-100)
 \put(10,0){$ \left( \ba{cccc|ccccccccc}
 \bt&\bt&&& \bt&.&\bt&&&&&&\\
 \bt&\bt&&& \bt&.&\bt&&&&&&\\
 \bt&\bt&\bt&\bt&  \bt&.&\bt&\bt&.&\bt&\bt&.&\bt\\
 \bt&\bt&\bt&\bt&  \bt&.&\bt&\bt&.&\bt&\bt&.&\bt\\
   \hline
 \bt&\bt&&&  \bt&.&\bt &&&&&&\\ .&.&&& .&.&.&&&&&&\\ \bt&\bt&&&
 \bt&.&\bt &&&&&&\\ \bt&\bt&&& \bt&.&\bt&\bt&.&\bt &&&\\ .&.&&&
 .&.&.&.&.&. &&&\\ \bt&\bt&&& \bt&.&\bt&\bt&.&\bt &&&\\
 \bt&\bt&\bt&\bt&  \bt&.&\bt&\bt&.&\bt&\bt&.&\bt\\ \bt&\bt&\bt&\bt&
 \bt&.&\bt&\bt&.&\bt&\bt&.&\bt\\ \bt&\bt&\bt&\bt&
 \bt&.&\bt&\bt&.&\bt&\bt&.&\bt
 \end{array} \right)$}
  \put(110,10) {$
 \left. \phantom{\begin{array}{c} \bt\\ .
 \\ \bt \end{array} } \right\}{\scriptscriptstyle p} $}
 \put(210,-70) {$ \left. \phantom{\begin{array}{c}  \\
 \\ \end{array} } \right\}{\scriptscriptstyle q} $} \put(155,-30) {$
 \left. \phantom{\begin{array}{c} \bt \\ . \\ \bt
 \end{array} } \right\}{\scriptscriptstyle N-p-q} $}
 \end{picture}
 \ee

A CR-analytic field defined on real $(N,p,q)$ harmonic superspace,
when continued to complexified superspace, will be a holomorphic
field defined on the above $(N,p,q)$ analytic superspace.

 \subsection{Parabolic induction}

If $G$ is a Lie group and $H$ a subgroup, representations of $G$
can be induced from representations of $H$ by considering
$H$-vector bundles over $H\bsh G$. Sections of such bundles are, as
we mentioned previously, equivalent to equivariant maps
$F:G\rightarrow V$, where $V$ is the representation space of $H$
with representation $R$, say. Thus we have (for $u\in G$)

\be
F(hu)=R(h)F(u) \ee

The induced representation itself is given by $F\mapsto g\cdot F,\ g\in
G$ where

\be
(g\cdot F)(u)=F(ug) \ee

It is easy to see that this does define a (left) representation of
$G$ carried by the space of sections of the vector bundle in
question. Parabolic induction refers to the case where the subgroup
is parabolic. In this case, if the representation $R$ is
irreducible then the representation of $G$ constructed in this
manner is also irreducible. (For a  detailed discussion of
parabolic subgroups and flag manifolds in the context of twistor
theory see \cite{be}.)

 Let G be a complex, simple Lie group and let $\gg$ be its Lie
 algebra. A Borel subalgebra $\gb$ is a maximal solvable subalgebra of
 $\gg$ and a parabolic subalgebra is one which contains a Borel
 subalgebra. For the case of $\gs\gl(N)$ we may take the Borel
 subalgebra to be the algebra of all lower triangular matrices (with
 non-zero entries on the diagonal allowed), and a parabolic subalgebra
 $\gp$ is one which is block lower triangular. Thus it consists of
 matrices of the form:

 \be
 \begin{picture}(300,200)(20,-100)
 \put(10,0){$\left(\ba{cccccccccc}
  \bullet&\bullet& &&& &&&&    \\
  \bullet&\bullet&&&&  &&&&\\
  \bullet&\bullet&\bullet&\bullet& & &&&&\\
  \bullet&\bullet&\bullet&\bullet& & &&&&\\
  \bt&\bt&\bt&\bt&\bt&\bt&\bt& &&\\
  \bt&\bt&\bt&\bt&\bt&\bt&\bt& &&\\
  \bt&\bt&\bt&\bt&\bt&\bt&\bt& &&\\
  &&&&&&&& .& \\
  &&&&&&&&&.
  \ea\right)$}
  \put(20,45) {$
 \left.\phantom{\ba{cc}
  \bullet&\bullet\\
  \bullet&\bullet\ea}\right\}
{\scriptscriptstyle k_1} $}
  \put(20,35) {$
 \left.\phantom{\ba{cccc}
  \bullet&\bullet&\bt&\bt\\
  \bullet&\bullet&\bt&\bt\\
  \bullet&\bullet&\bt&\bt\\
  \bullet&\bullet&\bt&\bt
 \ea}\right\}
{\scriptscriptstyle k_2} $}
  \put(20,15) {$
 \left.\phantom{\ba{ccccccc}
  \bullet&\bullet&\bt&\bt&\bt&\bt&\bt\\
 \bullet&\bullet&\bt&\bt&\bt&\bt&\bt\\
 \bullet&\bullet&\bt&\bt&\bt&\bt&\bt\\
 \bullet&\bullet&\bt&\bt&\bt&\bt&\bt\\
 \bullet&\bullet&\bt&\bt&\bt&\bt&\bt\\
 \bullet&\bullet&\bt&\bt&\bt&\bt&\bt\\
 \bullet&\bullet&\bt&\bt&\bt&\bt&\bt\ea}
  \right\}{\scriptscriptstyle k_3} $}
\end{picture}
 \ee

 where, if the squares have sides $k_1,k_2,\ldots k_{\ell}$, the
 parabolic specifies the flag manifold $\bbF_{\bfk}(N)$, where $\bfk=(k_1,k_2,\ldots k_{\ell})$. The parabolic
 $\gp$ can be represented by placing a cross on each of the nodes
 $k_1,k_2,\ldots k_{\ell}$ of the Dynkin diagram for $\gs\gl(N)$. For
 example, the flag manifold $\bbF_{1,3}(4)$, as well as the
 corresponding parabolic subalgebra, is represented by the $\gs\gl(4)$
 Dynkin diagram
 \begin{picture}(30,10)(0,0) \put(0,0){\makebox[0pt][l]{$\xz
 \bt\xz$} \rule[.5ex]{1.8em}{.1ex} }
 \end {picture}. Thus we have two constructions of flag manifolds -
 as homogeneous spaces of $SU(N)$ with isotropy groups of the type
 discussed previously, which makes it clear that they are
 compact, and as homogeneous spaces of $SL(N)$ defined by parabolic
 isotropy groups, which makes it clear that they are complex
 manifolds. As we mentioned previously, the Borel subgroup defines
 the space of flags
 of type $\bfk=(1,2,3,\ldots (N-1))$; it is the same space as the
 homogeneous space of $SU(N)$ which has the maximal torus as its
 isotropy group.

 The sub-algebra of $\gp$ consisting of the blocks along the diagonal
 is called the Levi sub-algebra $\gl$ and we have $\gp=\gl\oplus\gu$
 where $\gu$ consists of the remaining matrices; clearly $\gu$ is
 nilpotent, $\gu_k=0$ for some $k$ where $\gu_k$ is defined
 iteratively by $\gu_k=[\gu_{k-1},\gu_{k-1}];\ \gu_0=\gu$. For a
 general flag manifold $\bbF_{\bfk}(N)$, the Levi subalgebra is

 \be
 \gl=\gs\gl(k_1)\oplus \gs\gl(k_2-k_1)\oplus
 \ldots\oplus\gs\gl(N-k_{\ell})\oplus \bbC^{\ell} \ee

 The finite-dimensional irreducible representations of $\gs\gl(N)$ are
 specified by highest weights $\l$. These representations can
 alternatively be constructed by parabolic induction. In fact they act
 on holomorphic sections of homogeneous
 vector bundles over the flag manifolds $P\bsh SL(N)$. This procedure
 can be described explicitly. Choose a representation with highest
 weight $\l$ and fix a parabolic $\gp$. Any weight for $\gs\gl(N)$ is
 also a weight for any $\gp$ and so highest weights determine
 irreducible representations of $\gp$ in some vector space $V_{\l}$. In
 fact, since $\gu$ is nilpotent, it acts trivially on such a
 representation, so that we actually have a representation of the Levi
 subalgebra of $\gp$ on $V_{\l}$. We then construct the vector bundle
 $\cV_{\l}$
 over $P\bsh SL(N)$, i.e. the bundle whose standard fibre is the
 $P$-representation space $V_{\l}$. A simple application of the
 Bott-Borel-Weil theorem gives the desired result: the space of
 holomorphic sections of $\cV_{\l}$ is isomorphic to the representation
 space of $\gs\gl(N)$ determined by the highest weight $\l$. Note that
 a given representation can be presented in this fashion in as many
 different ways as there are parabolic subalgebras of $\gs\gl(N)$. ( A  discussion of this theorem in the context of twistor theory is given in \cite{be}).

 This discussion can be made more concrete as follows: a holomorphic
 section, $f$, of the vector bundle $\cV_{\l}$ is the same thing as a
 holomorphic map, $F$, from $SL(N)$ to the representation space $V$
 which is equivariant with respect to the subgroup $P$:

 \be
 F(p u)=R(p)F(u)
 \ee

 where $R$ denotes the representation in question (and is only
 non-trivial for the Levi subgroup $L$), and where $p\in P,u\in SL(N)$. If the
 representation in
 question has Dynkin labels $(a_1,\ldots,a_{N-1})$, it can be
 represented by a Young tableau with $m_l$ boxes in the $l$th row,
 $l=1,\ldots ,N-1$ with

 \be
 m_l=\sum_{k=l}^{N-1} a_k
 \ee

 Denote an element of $SL(N)$ by $u_I{}^i$. The
 index $I$ splits under the Levi sub-algebra as $I=(r_1,r_2,\ldots
 r_{\ell+1})$, where $r_i$ runs from $k_{i-1}$ to $k_i$, with $k_0=1$
 and $k_{\ell+1}=N$. The desired holomorphic section corresponding to
 the representation with Dynkin labels is schematically of the form

 \be
 F(u)\sim(u_1)^{m_1+m_2+\ldots m_{k_1}}(u_2)^{m_{k_1+1}+\ldots m_{k_2}}
 \ldots (u_{\ell+1})^{m_{k_{\ell}}\ldots m_{N-1}}\hat F
 \ee

 where $u_1$ denotes the $k_1\xz N$ matrix $u_{r_1}{}^i$, etc. and
 where $\hat F$ on the right-hand side denotes the SL(N) tensor with a
 total of $m=\sum_{k=1}^{N-1}ka_k$ (subscript) indices.

 \subsection{Flag supermanifolds}

 A super-flag in a supervector space $\bbC^{m|n}$ is a nested
 sequence of sub-supervector spaces $V_1\subset V_2 \ldots
 V_{\ell}\subset \bbC^{m|n}$ where the super-dimension of $V_i$ is
 $K_i=(k_i|\k_i)$, $k_i(\k_i)$ being the dimensions of the even (odd)
 subspaces of $V_i$ respectively, with $K_i<K_{j}$ for $i<j$.
 This notation means that $k_i\leq k_j,\,  \k_i\leq \k_j$, but
 with equality
 for both even and odd indices prohibited. The space of such
 super-flags will be denoted $\bbF_{\bfK}(m|n)$, where
 $\bfK=(K_1,\ldots K_{\ell})$ \cite{manin}. In the present paper we
 are interested
 in super-flags in $\bbC^{4|N}$, and they are determined by various
 parabolic subgroups of $SL(4|N)$. In general, there are three main
 classes of such spaces according to whether the body is  spacetime
 times an internal flag manifold, a twistor space (in the usual
 sense), or a combination of both, i.e. a twistor space times an
 internal flag manifold. We shall be interested in the first of these
 types of space since they correspond to (complexified) harmonic
 superspaces. Moreover, we shall be interested in superspaces which
 correspond to G-analyticity of type $(p,q)$. In the complex setting
 this means that we want to restrict our attention to spaces which
 have $(N-p)$ two-component undotted spinor coordinates and $(N-q)$
 two-component dotted spinor coordinates. The relevant  flag
 supermanifolds are thus
 given by $\bfK=(2|p,2|p+k_1,\ldots 2|p+k_{\ell},2|(N-q))$. The body
 of such a supermanifold is locally the product of spacetime with the
 internal flag manifold $\bbF_{p,p+k_1,\ldots, p+k_{\ell}, N-q}(N)$.
 Such supermanifolds are generalisations of the $(N,p,q)$ analytic
 superspaces
 introduced in \cite{hh2}. The latter are the flag supermanifolds
 $\bbF_{2|p,2|N-q}(4|N)$; they are the spaces with the smallest
 internal manifolds for a given choice of $(p,q)$. The $(N,p,q)$
 harmonic superspaces have internal flag manifolds $\bbF_{p,N-q}(N)$
 with associated Levi sub-algebras
 $\gs\gl(p)\oplus\gs\gl(q)\oplus\gs\gl(r)\oplus\bbC^2$, where
 $r=N-(p+q)$. The generalised $(N,p,q)$ spaces differ in that a
 non-trivial internal space is allowed corresponding to the central
 $r\xz r$ sector of the internal symmetry group.

 An important consequence of these definitions is that not all
 possible internal flag manifolds are compatible with a given $(p,q)$
 G-analyticity and superconformal symmetry. For example, if either $p$
 or $q$, or both, is bigger than 1, then the internal parabolic cannot
 be taken to be the Borel subalgebra. \footnote{This would seeem to
 contradict refs \cite{afsz,fs1}. However, the harmonic superfields
 for the underlying massless supermultiplets used by these authors
 satisfy further harmonic anti-analytic constraints which means that
 they actually live on $(N,p,q)$ spaces; in fact, they are the same fields as
 those of \cite{hh2}.} This may be clearly seen from
 the following pictorial representation of the parabolic subalgebra of
 $\gs\gl(4|N)$ for $(N,p,q)$ analytic superspace:

 \be
 \begin{picture}(300,200)(20,-100)
 \put(10,0){$ \left( \ba{cccc|ccccccccc}
 \bt&\bt&&& \bt&.&\bt&&&&&&\\
 \bt&\bt&&& \bt&.&\bt&&&&&&\\
 \bt&\bt&\bt&\bt&  \bt&.&\bt&\bt&.&\bt&\bt&.&\bt\\
 \bt&\bt&\bt&\bt&  \bt&.&\bt&\bt&.&\bt&\bt&.&\bt\\
   \hline
 \bt&\bt&&&  \bt&.&\bt &&&&&&\\ .&.&&& .&.&.&&&&&&\\ \bt&\bt&&&
 \bt&.&\bt &&&&&&\\ \bt&\bt&&& \bt&.&\bt&\bt&.&\bt &&&\\ .&.&&&
 .&.&.&.&.&. &&&\\ \bt&\bt&&& \bt&.&\bt&\bt&.&\bt &&&\\
 \bt&\bt&\bt&\bt&  \bt&.&\bt&\bt&.&\bt&\bt&.&\bt\\ \bt&\bt&\bt&\bt&
 \bt&.&\bt&\bt&.&\bt&\bt&.&\bt\\ \bt&\bt&\bt&\bt&
 \bt&.&\bt&\bt&.&\bt&\bt&.&\bt
 \end{array} \right)$}
  \put(110,10) {$
 \left. \phantom{\begin{array}{c} \bt\\ .
 \\ \bt \end{array} } \right\}{\scriptscriptstyle p} $}
 \put(210,-70) {$ \left. \phantom{\begin{array}{c}  \\
 \\ \end{array} } \right\}{\scriptscriptstyle q} $} \put(155,-30) {$
 \left. \phantom{\begin{array}{c} \bt \\ . \\ \bt
 \end{array} } \right\}{\scriptscriptstyle N-p-q} $}
 \end{picture}
 \ee

 Multiplication of odd elements of the algebra
 (i.e. the bottom left times the top right) clearly
 generates the $p\xz p$ and $q\xz q$ parts of the internal algebra
 which must therefore  be filled with bullets. The remaining central
 part of the 
 algebra can be completed in many different ways so that an internal
 parabolic subalgebra is obtained; we have depicted the simplest
 example here corresponding to the flag space $\bbF_{p,N-q}(N)$.

 The group $SL(4|N)$ acts to the left on $\bbC^{4|N}$ which may be
 considered as $N$-extended non-projective super twistor space \cite{f}. To
 exhibit the flag nature of the above superspaces more clearly it is
 convenient to make a change of basis of $\bbC^{4|N}$. If a
 supertwistor $Z$ is written as

 \be
 Z_A=\left(\ba{c} z^{\a} \\z_{\adt} \\ \z_i \ea\right)
 \ee

 then the required change of basis swaps $z_{\adt}$ and $\z_i$.

 \be
 Z_A\rightarrow \left(\ba{c} z^{\a} \\\z_i \\ z_{\adt} \ea\right)
 \ee

 In this basis the above parabolic takes the following form:

 \be
 \begin{picture}(300,200)(20,-100)
 \put(10,0){ $ \left( \ba{cc|ccccccccc|cc}
\bt &\bt &\bt &. &\bt&&&&&&&&
\\ \bt &\bt &\bt &. &\bt&&&&&&&&\\ \hline \bt &\bt &\bt &.
 &\bt&&&&&&&&\\ . &. &. &. &. &&&&&&&&\\ \bt &\bt &\bt &.
 &\bt&&&&&&&&\\ \bt &\bt &\bt &. &\bt &\bt &. &\bt &&&&&\\
 .&.&.&.&.&.&.&.&&&&&\\ \bt &\bt &\bt &. &\bt &\bt &. &\bt &&&&&\\ \bt
 &\bt &\bt &. &\bt &\bt &. &\bt&\bt&.&\bt&\bt&\bt\\
 .&.&.&.&.&.&.&.&.&.&.&.&.
\\ \bt &\bt &\bt &. &\bt &\bt &.
 &\bt&\bt&.&\bt&\bt&\bt\\ \hline
\bt &\bt &\bt &. &\bt &\bt &.
 &\bt&\bt&.&\bt&\bt&\bt\\
\bt &\bt &\bt &. &\bt &\bt &.
 &\bt&\bt&.&\bt&\bt&\bt\\ \ea \right) $ }
\put(80,40) {$
 \left. \phantom{\begin{array}{c} \bt\\ .
 \\ \times \end{array} } \right\}{\scriptscriptstyle p} $}
 \put(207,-40) {$ \left. \phantom{\begin{array}{c} \times \\ . \\ \bt
 \\ \end{array} } \right\}{\scriptscriptstyle q} $} \put(120,0) {$
 \left. \phantom{\begin{array}{c} \times \\ . \\ \bt
 \end{array} } \right\}{\scriptscriptstyle N-p-q} $}
 \end{picture}
 \end{equation}

 This brings it to block lower-triangular form as in the bosonic case.
 \footnote{In this basis the Borel subalgebra consists of the lower
 triangular matrices; note that it is possible to have inequivalent
 Borel subalgebras in the super case.}
 The Levi subalgebra for this parabolic is

 \be
 \gl=\gs(\gg\gl(2|p)\oplus\gg\gl(2|q)\oplus \gg\gl(r))
 \ee

 provided that $N\neq 4$. For non-exceptional cases, i.e. when neither
 $p$ nor $q$ equals $2$ and when $r\neq 0$, this algebra is isomorphic
 to $\gs\gl(2|p)\oplus\gs\gl(2|q)\oplus \gs\gl(r)\oplus \bbC^2$. We
 shall discuss the exceptional cases in the next section in more
 detail. In the case of generalised $(N,p,q)$ spaces, the central
 $\gs\gl(r)$ summand is replaced by an appropriate parabolic
 subalgebra.


\section{Short representations}


We can apply the method of parabolic induction straightforwardly in
the supersymmetric case to obtain representations of $SL(4|N)$ as
holomorphic sections of homogeneous vector bundles over flag
supermanifolds defined by parabolic subgroups of the type we have
discussed above. The short representations are characterised by
being short multiplets and thus having lower component spins than
unconstrained superfields on Minkowski superspace. Such
representations act naturally on superfields defined on analytic
superspaces since they have fewer odd coordinates than Minkowski
superspace.

The superfields carrying the representations should transform under
irreducible representations of Levi subalgebras of the form
$\gl=\gs(\gg\gl(2|p)\oplus\gg\gl(2|q)\oplus \gg\gl(r))$; however,
in order to ensure that the representations are indeed short these
superfields must not carry any spacetime indices. They must
therefore transform trivially under any supergroup factors of the
Levi subgroup. In the generic case this means that they transform
only under $\gs\gl(r)\oplus\bbC^2$, where we recall that
$r=N-(p+q)$. Another point to be borne in mind is that the same
representation can be realised on different superspaces
corresponding to different parabolic subgroups; in particular, for
a given representation there will be a class of ''maximally
efficient'' realisations, by which we mean that the shortening is
the largest possible, i.e. $p$ and $q$ take their largest possible
values for the given representation. There may be several
representations of this type due to the fact that one is free to
choose the central factor of the internal parabolic in different
ways.

In order to keep matters as simple as possible, we shall
concentrate for the time being on $(N,p,q)$ superspaces which have
internal flag manifolds $\bbF_{p,N-q}(N)$. The representations to
be studied can then be represented by modified Dynkin diagrams of
the following type:

\be
\begin{picture}(300,30)
\put(-20,0){\makebox[0pt][l]{$\bt\hspace{3em}\bt\hspace{3em}\xz\hspace{3em}
\bt\hspace{3em}\bt\hspace{3em}\bt\hspace{3em}\xz\hspace{3em}\bt\hspace{3em}
\bt$} \rule[.5ex]{30em}{.1ex} } \put(-20,10){$\tiny  0
\hspace{4.8em} 0\hspace{4.8em}a_{p}
\hspace{4.5em}a_{p+1}\hspace{5.5em}\ldots
\hspace{4.5em}a_{N-q}\hspace{4em} 0 \hspace{4.8em} 0 $}
\end {picture}
\ee

For the reasons discussed above the first $(p-1)$ and the last $(q-1)$ Dynkin labels must
vanish, leaving $(r-1)$ labels to specify the representation of the
central $\gs\gl(r)$ and two further labels which specify the
charges. The Young tableau for this representation is

\be
\setlength{\unitlength}{.3mm}
\begin{picture}(500,200)(0,0)
\put(400,140){\line(0,1){60}} \put(380,120){\line(0,1){80}}
\put(320,120){\line(0,1){80}} \put(200,60){\line(0,1){140}}
\put(20,180){\framebox(400,20)} \put(20,160){\framebox(400,20)}
\put(20,140){\framebox(400,20)}
\put(20,120){\framebox(360,20)}\put(395,120){$a_p$}
\put(20,100){\framebox(300,20)}\put(340,100){$a_{p+1}$}
\put(20,80){\framebox(280,20)} \put(20,60){\framebox(220,20)}
\put(20,40){\framebox(180,20)} \put(0,165){${\scriptscriptstyle
p}\left\{\phantom{\ba{c}\\ \\  \\
\\\ea} \right.$}
\put(0,85){${\scriptscriptstyle r} \left\{\phantom{\ba{c}\\ \\ \\
\\
\\ \\\ea} \right.$} \put(20,20){$\hspace{5em} a_{p+r}$}
\end{picture}
\la{tab}
\ee

From a purely algebraic point of view representations of
superconformal groups in four dimensions are determined by the
following set of labels: $(J_1,J_2,L,R;a_1,a_2,\ldots a_{N-1})$,
where $J_1,J_2$ are spins, $L,R$ are charges corresponding to
dilations and R-symmetry respectively while the $a's$ are
$\gs\gl(N)$ Dynkin labels (see, e.g., \cite{fz,afsz} for reviews). For representations of the above type
$J_1=J_2=0$, as are the Dynkin labels $a_1,\ldots a_{p-1};
a_{N-q+1}\ldots a_{N-1}$, while the $L$ and $R$ charges are, except
when either $p$ or $q=0, N\neq 4$ determined by these labels. The
representations that occur belong to series C with $J_1=J_2=0$ for
which one has

\be
L=m_1;\qquad R={2m\over N}-m_1 
\ee

where

\be
m:=\sum_{k=1}^{N-1}k a_k;\qquad m_1:=\sum_{k=1}^{N-1}a_k \ee

Note that $m$ is the total number of boxes in the $\gs\gl(N)$ Young
tableau, while $m_1$ is the number of boxes in the first row. When
$q=0, N\neq 4$, one has the weaker condition

\be
L-R=2m_1-{2m\over N} 
\la{L-R}
\ee

while if $p=0,N\neq4$, one finds

\be
L+R={2m\over N} 
\la{L+R}\ee

Both of these conditions are compatible with either series C or
series B. In the latter case one has, for zero spins, $L\geq m_1
+1$, together with either \eq{L-R} or \eq{L+R}. In these cases one more
label ($L$ or $R$) has to be specified in order to fix the
representation. For $N=4$ the same formulae hold, but now $R=0$ as
well. In particular, this means that the representations are fully
determined by the Dynkin labels even if $p$ or $q=0$. It should be noted that there is a second possibility for $N=4$ when the group is $SL(4|4)$ and not $PSL(4|4)$; this case is the same as general $N$.

To demonstrate the above explicitly, we first note that, in the
basis in which we are working, the matrices representing $L$ and
$R$ are

\be
L={1\over2}\left(\ba{c|ccc|c}
 -1_2&& && \\
 \hline
  &&& & \\
  &&\phantom{{4\over N} 1_N}&&\\
  &&&&\\
\hline
 &&&&1_2 \ea
\right) \ee

and

\be
R={1\over2} \left(\ba{c|ccc|c}
 1_2&& & \\
\hline &&&&\\ &&{4\over N}1_N&&\\ &&&&\\ \hline
 &&&& 1_2
 \ea \right)
 \ee

As we have mentioned, for most of these representations, fixing
$(p,q)$ and the Dynkin labels determines the values of $L$ and $R$
as well. The key point is that, as we remarked above, factors such
as $\gs\gl(2|p)$ must be represented trivially in order for the
fields we are interested in to have $J_1=J_2=0$. To put it another
way, the sub-algebra $\gg\gl(2|p)$ is represented by its
supertrace. This means that any element with vanishing supertrace
becomes zero when acting on fields of interest. In particular,

\be
\left(\ba{c|ccc|c} {1\over2}1_2 & & & &\\ \hline & {1\over p}1_p
&&&\\ &&&&\\ &&&&\phantom{{1\over2}1_2}\\ \hline &&&& \ea\right)
\rightarrow 0 
\la{1p}\ee

and, similarly

\be
\left(\ba{c|ccc|c}
 \phantom{{1\over2}1_2}& & & &\\
 \hline
&&&&\\ &&&&\\ &  &&{1\over q}1_q&\\ \hline
 &&&&{1\over2}1_2
 \ea\right) \rightarrow 0
\la{1q}
\ee

The operator $1_p$ on the representation in question counts the
total number of boxes in the first $p$ rows, of the Young tableau
\eq{tab}, while the operator $1_N$ counts the total number of boxes in
the  diagram. If we denote the field by $A$, then $1_p A=pm_1 A$,
$1_N A=m A$; since the tableau \eq{tab} has boxes only in the first $N-q$ rows it follows that $1_q A=0$.

It might be thought that when either $p$ or $q$ equals $2$ there
could be an extra charge due to the fact that $\gs\gl(n|n)$ is not
simple since the unit matrix has vanishing supertrace. However, for
vector representation spaces which do not carry any indices for this supergroup,
one can show that the unit matrix must be represented by zero
(because the anticommutator of two odd generators includes it) and
this means that the above rules for evaluating $L$ and $R$ apply in
this case, too.

In $N=4$ the superconformal group can be taken to be either
$SL(4|4)$ or $PSL(4|4)$. The latter group is simple and is the one
which is relevant for  SYM, while the former is needed for certain
representations such as the spin $3/2$ supermultiplet. In the
second case case it is still easier to work with $SL(4|4)$ and
impose the constraint $R=0$ to get to $PSL(4|4)$. Note that in
either case R-symmetry does not act on the coordinates.

We shall now go through the various cases that can arise,
explicitly evaluating $L$ and $R$ for the representations in
question:

\vskip .5cm

{\bf 1. The generic case: \boldmath $p,q,r\neq 0$}

In this case, the residual symmetry algebra acting on the representation space of the isotropy group $V$ is
$\gs\gl(r)\oplus\bbC^2$. Using the above rules for evaluating the
unit matrices $1_N,1_p,1_q$, we find, as expected,
\be
L=m_1;\qquad R={2m\over N}-m_1 \ee

The representation belongs to series C.

\vskip .5cm

{\bf 2. \boldmath $p,q\neq 0;r=0$}

In this case the residual symmetry algebra is simply $\bbC$ and the
Dynkin diagram reduces to

\be
\begin{picture}(300,30)
\put(-20,0){\makebox[0pt][l]{$\bt\hspace{3em}\bt\hspace{3em}\xz\hspace{3em}
\bt\hspace{3em}\bt\hspace{3em}\bt\hspace{3em}\bt$}
\rule[.5ex]{22.2em}{.1ex} } \put(-20,10){$\tiny 0 \hspace{4.8em}
0\hspace{4.8em}a_{p} \hspace{4.5em}0\hspace{6.7em}\ldots
\hspace{7em} 0 $}
\end {picture}
\ee

so that all the labels are zero except for $a_p$. $L$ and $R$ are
given by the same formulae with $m=pa_p;\,m_1=a_p$. The Young
tableau is

\be
\setlength{\unitlength}{.3mm}
\begin{picture}(300,100)(0,100)
\put(40,140){\line(0,1){60}} \put(60,140){\line(0,1){60}}
\put(80,140){\line(0,1){60}} \put(200,140){\line(0,1){60}}
\put(20,180){\framebox(200,20)} \put(20,160){\framebox(200,20)}
\put(20,140){\framebox(200,20)} \put(0,165){${\scriptscriptstyle
p}\left\{\phantom{\ba{c}\\ \\  \\
\\ \ea} \right.$}
\put(100,170){. . . } \put(100,120){\vector(-1,0){80}}
\put(140,120){\vector(1,0){80}} \put(120,115){$a_p$}
\end{picture}
\ee

\vskip .5cm

{\bf 3. \boldmath $p\neq 0,q=0$}

The Dynkin diagram now has the form:

\be
\begin{picture}(300,30)
\put(-20,0){\makebox[0pt][l]{$\bt\hspace{3em}\bt\hspace{3em}\xz\hspace{3em}
\bt\hspace{3em}\bt\hspace{3em}\bt\hspace{3em}\bt$}
\rule[.5ex]{22.2em}{.1ex} } \put(-20,10){$\tiny 0 \hspace{4.8em}
0\hspace{4.8em}a_p \hspace{4.5em}a_{p+1}\hspace{5em}\ldots
\hspace{6em} a_{N-1} $}
\end {picture}
\ee

In this case the Levi subalgebra is $s(\gg\gl(2|p)\oplus
\gg\gl(N-p)\oplus\gg\gl(2))$, where the last term arises from the
fact that $q=0$. In this case the constraint \eq{1q} is lost, and one
therefore only finds that

\be
L-R=2m_1-{2m\over N} \ee

This is compatible with both series B and C representations. \vskip
.5cm

{\bf 4. \boldmath $p=0,q\neq0$}

The Dynkin diagram now has the form:

\be
\begin{picture}(300,30)
\put(-20,0){\makebox[0pt][l]{$\bt\hspace{3em}\bt\hspace{3em}\bt\hspace{3em}
\bt\hspace{3em}\xz\hspace{3em}\bt\hspace{3em}\bt$}
\rule[.5ex]{22.2em}{.1ex} } \put(-20,10){$\tiny a_1 \hspace{4em}
a_2\hspace{5.7em}\ldots \hspace{5.7em}a_{N-q}\hspace{4.5em}0
\hspace{4.5em} 0 $}
\end {picture}
\ee

In this case the Levi subalgebra is $s(\gg\gl(2|q)\oplus
\gg\gl(N-q)\oplus\gg\gl(2))$, where the last term now arises
because $p=0$. Again the constraint \eq{1p} is lost, and one
therefore only finds that

\be
L+R={2m\over N} \ee

This is also compatible with series B and C.

For both $(p,q)=(p,0)$ and $(p,q)=(0,q)$ there is the possibility
of having superfields with all Dynkin labels equal to zero. These
fields just behave as holomorphic functions with regard to the
internal space and  therefore do not depend on the coordinates of
the internal space by compactness. G-analyticity then implies that
they are chiral or anti-chiral scalar fields (for $(0,q)$ or
$(p,0)$ respectively). For $L=1$ such superfields are not
irreducible and satisfy an additional constraint, as we shall
discuss in the next section, but for other (higher) values of $L$
they are irreducible superconformal fields with $L=\pm R$.

\vskip .5cm The remaining examples all refer to  $N=4$ with the
group $PSL(4|4)$.

{\bf 5. \boldmath $(p,q)=(1,3)$}

This is a particular example of $(N,p,N-p)$ analytic superspace.
In principle, one would expect to have Dynkin diagrams of the form

\be
\begin{picture}(300,30)
\put(-20,0){\makebox[0pt][l]{$\xz\hspace{3em}\bt\hspace{3em} \bt$}
\rule[.5ex]{7.8em}{.1ex} } \put(-20,10){$\tiny  a_1 \hspace{4.8em}
0 \hspace{4.5em} 0 $}
\end {picture}
\ee

but the imposition of $R=0$ implies that $a_1=0$, so that there are
no short representations on this space.

\vskip .5cm

{\bf 6. \boldmath $(p,q)=(2,2)$}

The Dynkin diagram is:

\be
\begin{picture}(300,30)
\put(-20,0){\makebox[0pt][l]{$\bt\hspace{3em}\xz\hspace{3em} \bt$}
\rule[.5ex]{7.8em}{.1ex} } \put(-20,10){$\tiny  0 \hspace{4.8em}
a_{2} \hspace{4.5em} 0 $}
\end {picture}
\ee

and this is compatible with $R=0$.

\vskip .5cm

{\bf 7. \boldmath$(p,q)=(1,2)$}

We expect to have

\be
\begin{picture}(300,30)
\put(-20,0){\makebox[0pt][l]{$\xz\hspace{3em}\xz\hspace{3em} \bt$}
\rule[.5ex]{7.8em}{.1ex} } \put(-20,10){$\tiny  a_1 \hspace{4.8em}
a_{2} \hspace{4.5em} 0 $}
\end {picture}
\ee

but $R=0$ implies that $a_1=0$; hence there are no new $(1,2)$
representations, since  the ones that are allowed can be formulated
on $(2,2)$ superspace.

\vskip .5cm

{\bf 8. \boldmath $(p,q)=(1,1)$}

In this case we could expect to have arbitrary representations of
$\gs\gl(4)$,

\be
\begin{picture}(300,30)
\put(-20,0){\makebox[0pt][l]{$\xz\hspace{3em}\bt\hspace{3em} \xz$}
\rule[.5ex]{7.8em}{.1ex} } \put(-20,10){$\tiny  a_1 \hspace{4.8em}
a_{2} \hspace{4.5em} a_3 $}
\end {picture}
\ee

However, $R=0$ implies that $a_1=a_3$.

\vskip .5cm

{\bf 9. \boldmath$(p,q)=(p,0)$}

The possible diagrams are

\bea &&\begin{picture}(300,20)
\put(-20,0){\makebox[0pt][l]{$\xz\hspace{3em}\bt\hspace{3em} \bt$}
\rule[.5ex]{7.8em}{.1ex} } \put(-20,10){$\tiny  a_1 \hspace{4.8em}
a_{2} \hspace{4.5em} a_3$}
\end {picture} \nn\\
&&\begin{picture}(300,50)
\put(-20,0){\makebox[0pt][l]{$\bt\hspace{3em}\xz\hspace{3em} \bt$}
\rule[.5ex]{7.8em}{.1ex} } \put(-20,10){$\tiny  0 \hspace{4.8em}
a_{2} \hspace{4.5em} a_3$}
\end {picture}\nn\\
&&\begin{picture}(300,50)
\put(-20,0){\makebox[0pt][l]{$\bt\hspace{3em}\bt\hspace{3em} \xz$}
\rule[.5ex]{7.8em}{.1ex} } \put(-20,10){$\tiny  0 \hspace{4.8em} 0
\hspace{5em} a_3$}
\end {picture}
\eea

for $p=1,2,3$ respectively. In all cases we find that

\be
L=2m_1-{m\over2} \ee

because $R=0$ and $q=0$. This can be rewritten as

\be
L=m_1 + {1\over2}(a_1-a_3) \ee

Since the superfields under consideration are bosonic it follows
that $L$ must be integral, and hence  $a_1-a_3=2b$, so that

\be
L=m_1+b \geq m_1 +1\ {\rm for}\ b\geq 1 \ee

There are therefore no $p=3$ representations, while for $p=2$ only
$a_3=0$ is allowed. Finally, for $p=1$ we have $a_1-a_3=2b$ so
that, for $b\neq 0$ we have series B with $L=m_1 + b$, while if
$b=0$ we have series C with $a_1=a_3$.

{\bf 10. \boldmath $(p,q)=(0,q)$}

The diagrams are the same as in the previous case, but now they
correspond to $q=3,2,1$ respectively. In all cases we have

\be
L={m\over2}=m_1 +{1\over 2}(a_3-a_1) \ee

There are no representations for $q=3$, for $q=2$ we can only have
$a_2\neq 0$ and for $q=1$ we can have series B with $L=m_1+b$ and
$a_3-a_1=2b\geq 2$, or series C with $L=m_1$, $a_1=a_3$.

\vskip.5cm

We conclude this survey of the various short representations with a
brief discussion of the general case. For the generic case, we can
have representations of the following type:

\be
\begin{picture}(300,50)(0,-30)
\put(-40,0){\makebox[0pt][l]{$\bt\hspace{3em}\xz\hspace{3em}\bt\hspace{3em}
\xz\hspace{6em}\bt\hspace{3em}\xz\hspace{3em}\bt\hspace{3em}\xz
\hspace{3em}\bt$} \rule[.5ex]{33.6em}{.1ex} } \put(-40,10){$\tiny 0
\hspace{5em} 0\hspace{5em}a_s \hspace{5em}\cdot\ \cdot\
\cdot\hspace{3em}\hspace{3em}a_t\hspace{5em} 0 \hspace{5em} 0
\hspace{5em}0\hspace{5em}0 $} \put(-40,-20){$\tiny  \hspace{5.5em}
p\hspace{5em}\phantom{a_s} \hspace{4.5em}p+k_1
\hspace{11em}p+k_{\ell}\hspace{8em}N-q $}
\end {picture}
\ee

This diagram is read as follows: the crosses determine the internal
parabolic subalgebra of $\gs\gl(N)$, the corresponding internal
flag manifold being $\bbF_{p,p+k_1,\ldots p+{k_{\ell}},N-q}(N)$;
the two outer crosses, at the $p$th and $(N-q)$th nodes, determine
the explicitly realised G-analyticity to be of type $(p,q)$, and
the non-zero Dynkin labels are $a_s\ldots a_t$ where $s\geq p$ and
$t\leq N-q$. If $s=p$, $t=N-q$ the representation is maximally
efficient, but otherwise this is not the case. This holds if
neither $p$ nor $q$ is zero. The representation of the
superconformal group is fixed by the diagram since $L$ and $R$ can
be determined and the spins are zero. If either of $p$ or $q$, say
$q$, equals zero, then the only restriction is that the left-most
cross must be placed at node $p$, with $s\geq p$. As we have seen,
in this case the representation is only fully determined when an
extra label is specified unless $N=4$ and the group is $PSL(4|4)$
where the fact that $R=0$ implies that no extra label is required.

Finally, we note that any short representation can be realised on
$(1,0)$ superspace, while any representation from series $C$ can be
realised on $(1,1)$ superspace, although these realisations will
not be maximally efficient in general.


\section{Massless supermultiplets}


On-shell massless supermultiplets with maximal helicity $s$, where
$[\frac N2]\leq 2s < N$, ($[\frac N2]$ denotes the nearest
integer greater than or equal to $\frac N2$),  are described in (real) super Minkowski
space, $M$, by superfields $W$ which have $p=2s$ totally
antisymmetric internal indices and which satisfy \cite{s,hst}

\begin{eqnarray}
\bar D_{\adt}^i W_{j_1 \ldots j_{p}} & = &
\frac{p(-1)^{p-1}}{N-p+1} \delta_{[j_1}^i \bar D_{\adt}^k W_{j_2 \ldots
j_{p}] k} \nonumber \\ D_{\alpha i}W_{j_1 \ldots j_{p}} & = &
D_{\alpha [ i} W_{j_1 \ldots j_{p}]} \label{DW}
\end{eqnarray}

For each such superfield there is a conjugate superfield $\tilde
W_{i_1\ldots i_{N-p}}$ defined by

\be
\tilde W_{i_1\ldots i_{N-p}}={1\over p!}\vare_{i_1\ldots
i_{N-p}j_{N-p+1}\ldots j_N} \bar W^{j_{N-p+1}\ldots j_N} \ee

The conjugate superfield obeys similar constraints:

\begin{eqnarray}
\bar D_{\adt}^i \tilde W_{j_1 \ldots j_{N-p}} & = &
\frac{(N-p)(-1)^{N-p-1}}{p+1} \delta_{[j_1}^i \bar D_{\adt}^k \tilde
W_{j_2 \ldots j_{N-p}] k} \nonumber \\ D_{\alpha i}\tilde W_{j_1
\ldots j_{N-p}} & = & D_{\alpha [ i}\tilde W_{j_1 \ldots j_{N-p}]}
\label{eq:DW}
\end{eqnarray}

When $s=\frac{1}{4}N$, the multiplet is self-conjugate:

\begin{equation}
\bar{W}^{i_1 \ldots i_{p}} = \frac{1}{p!} \varepsilon^{i_1 \ldots
i_{p} j_1 \ldots j_{p}} W_{j_1 \ldots j_{p}}.
\end{equation}

For each $N$ there is therefore a range of scalar superfields $W_{i_1\ldots i_p}$
antisymmetric on $p$ internal indices with $p$ ranging from 1 to
$N-1$ all of which obey the constraints \eq{DW}. We can extend this to $N$, since for such a superfield the
constraints \eq{DW} imply that it is anti-chiral. Its conjugate has
no indices and is chiral. Such a chiral field describes an on-shell
massless supermultiplet (with maximum spin $N/2$) if it satisfies
the additional constraint \cite{fs1}

\be
D_{\a i} D^{\a}_j W=0
\la{chiraleom}
\ee

Note that this equation is only superconformal if the dilation
weight $L=1$; for other values of $L$ a chiral field is
superconformally irreducible and \eq{chiraleom} should not be
imposed.

The superfields $W_{i_i\ldots i_p}$ can be naturally described as
CR-analytic fields on $(N,p,N-p)$ harmonic superspaces \cite{hh2}.
On $M\xz SU(N)$ define the superfield $W^{(p)}$ by

\be
W^{(p)}=\frac{1}{p!}\vare^{r_1\dots r_p}u_{r_1}{}^{i_1}\dots
u_{r_p}{}^{i_p} W_{i_1\dots i_p}, \ee

then equation \eq{DW} implies that $W^{(p)}$ satisfies the
G-analyticity constraints

\bea D_{\a r} W^{(p)} &= & 0\\ \bar D_{\adt}^{r'}W^{(p)} &=&0 , \eea

and the H-analyticity constraint,

\be
D_r{}^{s'}W^{(p)}=0 \ee

In addition $W^{(p)}$ is clearly equivariant with respect to the
isotropy group $S(U(p)\xz U(N-p))$ of the internal flag manifold
$\bbF_{p,N-p}(N)$, which in this case is the Grassmannian of
$p$-planes in $\bbC^N$, $\bbG_p(N)$. Furthermore, an equivariant
field with the same charge as $W^{(p)}$ satisfying the above
analyticity constraints is equivalent to a superfield on super
Minkowski superspace satisfying \eq{DW}. We can apply this
construction to the entire family of scalar superfields with $p$
varying from $0$ to $N$, bearing in mind that the chiral fields
satisfy the additional constraint \eq{chiraleom}.

Under harmonic conjugation the superfield $W^{(p)}\mapsto \widetilde
{(W^{(p)})}=W^{(N-p)}$.

In complexified superspace, there are corresponding fields $W^{(p)}$
and $\tilde W^{(p)}$ defined on $(N,p,N-p)$ and $(N,N-p,p)$ analytic
superspace respectively. Furthermore, it is possible to extend
harmonic conjugation to this case straightforwardly. The Dynkin
diagram for $W^{(p)}$ is

\be
\begin{picture}(300,30)
\put(-20,0){\makebox[0pt][l]{$\bt\hspace{3em}\bt\hspace{3em}\bt\hspace{3em}
\bt\hspace{3em}\bt\hspace{3em}\xz\hspace{3em}\bt\hspace{3em}\bt\hspace{3em}
\bt$} \rule[.5ex]{30em}{.1ex} } \put(-20,10){$\tiny  0
\hspace{4.8em} 0\hspace{7em}\ldots \hspace{6.7em}0 \hspace{5em}
1\hspace{4.8em} 0\hspace{4em}\ldots\hspace{4em}0 $}
\end {picture}
\ee

where the single non-zero label is above the $p$th node, and thus
corresponds to the representation of $\gs\gl(N)$ on
$\wedge^P(\bbC^N)$. The diagram for $\tilde W^{(p)}$ is similar, but with
$p$ replaced by $N-p$.

Although $W^{(p)}$ provides the most efficient realisation of this
representation, the maximum spin for $W^{(p)}$, when $p\geq N/2$, is in fact only
$(0,\frac p2)$, and not $(\frac{N-p}{2},\frac p2)$; these
representations are therefore ultra-short. For $\tilde W^{(p)}$ the
maximum spin is $(\frac{p}{2},0)$. In the self-conjugate case, $W^{(p)}$ contains both $(0,\frac p2)$ and $(\frac p2,0)$.

There are many other ways of representing such multiplets which are
``less efficient'' in that the superspaces have more odd
coordinates. Following the discussion at the end of the previous
section we can simply place crosses where we like on the above
Dynkin diagram with the restrictions that the cross furthest to the
left must have node number $k$ less or equal to $p$ and the cross
furthest to the right must have node number $l$ at least equal to
$p$. The corresponding superspace will then have G-analyticity of
type $(k,l)$. For example, any such field (excluding $p=0,N$) can be realised with
$(1,1)$ G-analyticity on an internal flag defined by the Borel
subgroup (i.e. crosses at all nodes).

To illustrate this procedure we consider the $N=4$ Maxwell
supermultiplet which is represented on $N=4$ super Minkowski space
by the familiar self-conjugate Sohnius superfield $W_{ij}$
\cite{sohn}. The possible realisations of this superfield have
G-analyticities $(2,2),\ (1,1)$ and $(1,0)$. The $(2,2)$ superfield
is the one introduced in \cite{hh2}. The (1,1) superfield can be
written $W_{1r},\ r\in\{2,3\}$; in $(4,1,1)$ harmonic superspace it
obeys the constraints

\be
D_{\a 1}W_{1r}=\bar D_{\adt}^4 W_{1r}=0 \ee

Under harmonic conjugation we have

\be
W_{1r}\mapsto \tilde W^{4r}=\e^{rs}W_{1s}
\ee

The $(1,0)$ version is a superfield $W_{1r}$, $r\in\{2,3,4\}$
satisfying

\be
D_{\a 1}W_{1r}=0
\la{410}
\ee

while the $(0,1)$ version is a superfield $W_{RS},\ R\in{1,2,3}$ obeying

\be
\bar D_{\adt}^4 W_{RS}=0 \ee

Under harmonic conjugation we find

\be
W_{1r}\mapsto \tilde W^{4 R}={\frac 12}\e^{RST}W_{ST}
\ee

Due to self-conjugacy, the constraint \eq{410} together with
harmonic analyticity is sufficient to reproduce the standard
constraints obeyed by the Sohnius superfield.

The above superfields have been defined on $(4,1,1)$ and $(4,1,0)$
or $(4,0,1)$ superspaces which have the smallest possible internal
flags. It is possible to relax this; for example we could use the
maximal flag space determined by the Borel subalgebra. In the
$(1,1)$ case we would then split the indices as $I=1,2,3,4$; this
enables us to define the field $W_{12}$ which is $(1,1)$ G-analytic
and H-analytic on $\bbF_B$. Note that this field satisfies further first-order
differential constraints; it is actually $(2,2)$ G-analytic and is
also annihilated by $D_2{}^1$ and $D_4{}^3$


\section{Tensor products}


General short multiplets can be obtained as tensor products of the
massless supermultiplets in a straightforward fashion. If one
wishes to find such a short multiplet with $(p,q)$ analyticity the
optimal way to construct it from massless multiplets is to use
realisations of the latter with the same analyticity. One then
multiplies an appropriate number of massless multiplets together to
get the desired representation. Because the massless multiplets
will in general transform under non-trivial representations of the
internal symmetry algebra ($\gs\gl(r)$ in the case of $(N,p,q)$
superspace), it will be necessary to project out an irreducible
representation of this algebra in order to obtain the desired
irreducible short representation.

\subsection{Generating all short representations}

There is a very simple way of generating all possible short
representations from the set of available massless supermultiplets
for a given $N$. Consider first those belonging to series C. We
realise each of the superfields $W^{(p)}$ on $(N,1,1)$ superspace with
$\bbF_B$ as the internal flag space. Thus we have superfields $W_1,
W_{12}, W_{123}, \ldots W_{123\ldots N-1}$ all of which are
G-analytic with respect to $D_{\a 1}$ and $\bar D_{\adt}^N$. Let
$(a_1,\ldots a_{N-1})$ be the Dynkin labels of the representation
we are interested in and simply form the product

\be
A=\prod_{k=1}^{N-1}\left(W_{12\ldots k}\right)^{a_k} \la{prod}
\ee

The leading component of this superfield transforms according to
the irreducible representation of $SL(N)$ specified by the Dynkin
labels. The dilation weight is just equal to the sum of the Dynkin
labels, that is $m_1$, since each underlying field has weight 1.

This construction of a general representation also illustrates what
happens when the Dynkin labels are restricted. For example, suppose
that the first $p-1$ and the last $q-1$ labels are zero, then the
only superfields present in the above product will be $W_{12\ldots
p}, \ldots W_{12\ldots N-q}$. Each of these superfields is
G-analytic with respect to $D_{\a 1}\ldots D_{\a p}$ and $\bar
D_{\adt}^{N-q+1}\ldots \bar D_{\adt}^N$; furthermore, as well as
being H-analytic on $\bbF_B$ it satisfies further harmonic
anti-analyticity conditions and can therefore be thought of as a
superfield on $\bbF_{p,p+1,p+2,\ldots N-q-1,N-q}(N)$ superspace.

For series B representations we can work on $(N,0,1)$ superspace
again with $\bbF_B$ as the internal flag manifold. We use the same
family of superfields as before augmented with $W_o$ which
satisfies $\bar D_{\adt}^N W_o=0$ together with the additional
condition \eq{chiraleom}. We can then form the product superfield

\be
A'=(W_o)^b\prod_{k=1}^{N-1}\left(W_{12\ldots k}\right)^{a_k} \ee

where $b$ is a positive integer. The $L$ weight of this field is
$b$ plus the dilation weight of the previous example, so $L=m_1+b$.
Likewise the R-symmetry weight differs from the previous example by
$-b$, so $R=2m/N-(m_1+b)$ and therefore $L+R=2m/N$ as required. In
general this superfield will only be annihilated by $\bar
D_{\adt}^N$, but if the last $q-1$ Dynkin labels vanish, the
superfield will be defined on $(N,0,q)$ superspace.

If the field $A$ of \eq{prod} is multiplied by products of the
anti-chiral field with $L=1,R=1$ we get the B series of
representations on $(N,1,0)$ superspace with $L-R=2m_1-2m/N$.

\subsection{Products of a given massless multiplet}

In order to find the short representation composites that can be constructed using a given underlying massless supermultiplet one first of all computes the different Young tableau that can arise by multiplying together the tableaux for $k$ factors of $W$ and $l$ factors of $\tilde W$. For series C, these tableaux need to be subjected to the following restrictions: firstly, there should be $k+l$ boxes in the first row, secondly, there should be a total of $pk+(N-p)l$ boxes with at most $N-1$ in the first column and thirdly, a number of representaions will be absent due to symmetrisation. In the case of series B, one can either have columns with $N$ boxes, in which case after these have been discarded the remaining tableaux will clearly have fewer than $pk+(N-p)l$ boxes, or one is still restricted to at most $N-1$ boxes in a column, but with the constraint that there should be $k+l$ boxes in the first row dropped. The problem of restrictions due to symmetrisation can be overcome if one is allowed to use different copies of the same underlying multiplet.

As a simple example consider the $N=3$ Maxwell multiplet  which is
described by an on-shell superfield $W_{ij}$ of the above type. The
corresponding $W$ is CR-analytic on $(3,2,1)$ harmonic superspace,
and its conjugate $\tilde W$ is  CR-analytic on $(3,1,2)$ harmonic
superspace. Both of them can be realised as CR-analytic superfields
on $(3,1,1)$ superspace, with

\be
 W_{12}=u_1{}^i u_2{}^j W_{ij};\qquad \tilde W_1=u_1{}^i\tilde
W_i \ee

On $(3,2,1)$ superspace, the products of $W$ with itself have
maximum spins ${\frac 12,1}$, and corresponding tableaux
with 2 rows of the same length. The simplest of these
supermultiplets, $S$, say, has its leading component in the
6-dimensional representation of $SU(3)$. In order to construct the
energy-momentum tensor multiplet one has to take the product of $W$
and $\tilde W$ on $(3,1,1)$ superspace. This multiplet, $T$, say,
has maximum spin $(1,1)$ and its leading component is in the
$8$-dimensional representation of $SU(3)$. The multiplets $T$ and
$S$ are the $N=3$ ``components'' of the $N=4$ supercurrent. (See \cite{afsz} for more details of $N=3$ mulitplets.)

An interesting set of multiplets can be found for supergravity
theories with $N\geq 5$. The basic scalar superfield with
maximum spin 2 has four indices and is self-conjugate in $N=8$. The
analogue of the energy-momentum tensor in some senses for this
multiplet is the Bel-Robinson tensor ($C_{\a\b\c\d}\bar
C_{\adt\bdt\cdt\ddt} + \ldots$ in two-component spinor notation,
where $C$ is the Weyl spinor). This spin 4 field is the highest
component of the composite multiplet obtained by multiplying $W$
and its conjugate. This superfield lives naturally on $(N,N-4,N-4)$
superspace. If we square it, we obtain the three-loop supergravity
counterterm supermultiplet \cite{sg}. In $N=8$ this is simply $W^4$ since $W$
is self-conjugate \cite{hh2}.

\subsection{N=4 composites}

We conclude this discussion of composite operators with some comments on $N=4$ SYM since this 
is of most interest in the AdS/CFT context. The superconformal group in this case is $PSU(2,2|4)$ and these fields were discussed from a different perspective in \cite{afsz}.

We begin with Maxwell multiplets which we denote by $A,B,C$, etc. If
we regard these as superfields on $(4,2,2)$ superspace, the only
composites we can build are of the form $A^k$ with Dynkin labels
$(0,k,0)$ and highest spin $(1,1)$. On (1,1) superspace we can
generate all the series C representations using two fields
$A_{1r},B_{1r}$, where $r\in\{2,3\}$. The product $A^k B^l$, where
we may assume $k\geq l$, gives rise to a series of representations
with Dynkin labels $(n,k+l-2n,n)$ where $n=0,1,\ldots l$. The point
here is that products of $A$ with itself must be symmetric and can
therefore have tableaux with only two rows. To get three rows it is
necessary to introduce a second field $B$. Finally, to get
representations from series B one has to use a third field $C$ and
work on $(1,0)$ superspace. In this manner one can generate
tableaux with four rows, although one then discards the columns
which do have four rows.

In the non-Abelian theory the field strength superfield $W$ takes
its values in the Lie algebra of the gauge group which we shall
take to be $SU(N_c)$; it is not itself a CR-analytic superfield on
$(4,2,2)$ harmonic superspace, rather it is H-analytic and
covariantly G-analytic \cite{hh2}. However, gauge-invariant products of powers
of $W$ are CR-analytic and hence can be used to construct short
representations of the $N=4$ superconformal group. The simplest
series is the KK series $A_k:=\Tr(W^k)$; this series of operators
on $(4,2,2)$ space has maximum spin $(1,1)$ and includes the
supercurrent for $k=2$ \cite{hh2,hw}. To obtain other short representations from
series C we consider the field strength tensor as a $(4,1,1)$
superfield $W_{1r}, r\in\{2,3\}$. We can then form the double trace
operators $\Tr(W_{1r}^k)\Tr(W_{1s}^l)$, which of course decompose
into irreducible representations under $\gs\gl(2)$. Finally, we can
obtain series B representations by taking triple traces of a
similar type using the basic superfield $W_{1r},r\in\{2,3,4\}$
which is covariantly G-analytic on $(4,1,0)$ superspace. Each
factor in these multi-trace operators, if we assume the gauge-group
trace to be symmetrised, is one of the operators in the $A_k$
series, so that another way of obtaining these operators is as
products of these. For series C has to write these operators in
$(4,1,1)$ superspace, thus $A_k=\Tr(W^k)$ is replaced by $A_{r_1\ldots r_k}:=\Tr(W_{1r_1}\dots W_{1r_k})$. One
can then form the products $A_{r_1\ldots r_k} A_{s_1\ldots s_l}$
which can subsequently be decomposed into irreducible representations under
$SU(2)$. Similarly the triple trace operators can be viewed as
products of sets of three $A_k$ operators written as superfields on
$(4,1,0)$ superspace. The multi-trace operators are also discussed in \cite{afsz}.

It would seem that further gauge-invariant operators could be
constructed using the structure constants,
; for example, on $(1,0)$
superspace, there is an operator
$\vare^{rst}\tr(W_{1r}W_{1s}W_{1t})$ where the trace is no longer symmetrised.

\section{Superconformal transformations on harmonic superspace}

The short representations obtained by parabolic induction in
complex superspaces transform under the superconformal group in the
standard way. These transformations can then be adapted to real
spacetime. However, if one is interested in real harmonic
superspaces in the GIKOS formalism it is perhaps just as easy to
derive the explicit transformation rules by other means. This is
what we shall do in this section. Most of the results given here were obtained previously in \cite{hh1}, but the derivation is perhaps slightly simpler. We start from superconformal transformations on super Minkowski space, extend these in a simple way to $M\xz SU(N)$ and then make a final modification in order to preserve analyticity.

We recall that a superconformal transformation on super Minkowski
space is a diffeomorphism which preserves the odd tangent space
(spanned by the odd covariant derivatives). Dually, such a
transformation can be thought of as one which preserves the even
cotangent bundle spanned by the differential forms
$E^{\a\adt}:=dx^{\a\adt}+{i\over2}d\th^{\a i}\bar\th^{\adt}_i +
{i\over2}d\bar\th^{\adt}_i\th^{\a i}$. An infinitesimal such
transformation is generated by a superconformal Killing vector $V$
where

\be
V=F^{\a\adt}\del_{\a\adt} + f^{\a i}D_{\a i}-\bar f^{\adt}_i\bar
D^i_{\adt} \ee

with

\be
D_{\a i}F_{\b\bdt}=-i\vare_{\a\b}\bar f_{\bdt i} 
\la{scF}
\ee

so that

\be
f^{\a i}={i\over2}\bar D_{\adt}^i F^{\a\adt} \ee

Thus such a vector field is determined by the function $F^{\a\adt}$
obeying the above constraint. The Lie algebra can therefore be
presented in terms of $F$,

\be
[V_F,V_{F'}]=V_{[F,F']} \ee

where

\be
[F,F']^{\a\adt}:=F^{''\a\adt}=\left(F\cdot\del F'^{\a\adt}-if^{\a
i}\bar f^{'\adt}_i\right) - (F\leftrightarrow F') \ee

It is not difficult to verify that the superfield $F$ subject to
\eq{scF} does indeed specify a superconformal transformation. In $N=4$
there is an additional constraint, namely

\be
D\cdot f+\bar D\cdot \bar f;=D_{\a i}f^{\a i}+\bar D_{\adt}^i\bar
f^{\adt}_i=0 \la{n4constraint} \ee

For general $N$ one has the identity

\be
\del\cdot F-{2\over N}\left(D\cdot f-\bar D\cdot \bar f\right)=0
\ee

and we identify the dilation and R-symmetry parameters $L$ and $R$
by

\bea \del\cdot F&=&4L+\ldots \nn \\ D\cdot f+\bar D\cdot \bar
f&=&-2i(N-4)R,\ {\rm N\neq 4} \eea

The $N=4$ constraint \eq{n4constraint} has the geometrical consequence that the
$N=4$ chiral measure is superconformally invariant, since it
implies that $\del\cdot F -D\cdot f=0$.

The above transformation can be extended to a transformation of
$M\xz SU(N)$. The corresponding Killing vector in this case has the
form

\be
\cV=V_o + V_u \ee

where $V_o$ is a superconformal Killing vector on super Minkowski
space and

\be
V_u:=f_I{}^J D_J{}^I \ee

with

\be
f_I{}^J={1\over2}\left(D_{\a I} f^{\a J}-{1\over N}\d_I{}^J D_{\a
K} f^{\a K}\right) \ee

The parameter $f_{I}{}^J$ is the parameter superfield $f_i{}^j$ with indices converted to capital ones by means of $u$'s. The leading component of $f_i{}^j$ is the $\gs\gu(N)$ parameter. 

It is straightforward to verify that $\cV$ gives a
representation of the superconformal algebra.

If $A$ is an equivariant superfield on $M\xz SU(N)$ then it would
be natural to look for a transformation rule of the form $\d A=\cV
A$. This preserves H-analyticy, but not preserve G-analyticity.
This deficiency can be overcome by including an extra term in the
transformation rule. This term is constructed using the chiral
function $K$ defined by

\be
K={1\over4}\del\cdot F+{1\over 2(N-4)}\left( D\cdot f + \bar D\cdot
\bar f\right) \ee

which is normalised so that its leading component is $S-iR$. It is
also proportional to the transformation factor of the chiral
measure. A term of the form $(a K+ b\bar K)A$ can be added to the
transformation rule for $A$ because

\be
V_F K_{F'}-V_{F'}K_F=K_{[F,F']} \ee

Moreover, it is the only available function which preserves H-analyticity.

We consider a field $A$ which has $(p,q)$
G-analyticity and which is H-analytic in the sense of $(N,p,q)$
harmonic superspace. As before, we set $I=(r,r'',r')$. The field $A$ satisfies the G-analyticity
conditions

\be
D_{\a r}A=\bar D_{\adt}^{r'}A=0 \ee

as well as the H-analyticity conditions

\be
D_{r}{}^{s'}A=D_{r}{}^{s''}A=D_{r''}{}^{s'}A=0 \ee

$A$ cannot transform under $SU(p)$ (acting on $r$) or $SU(q)$
(acting on $r'$) because this is incompatible with G-analyticity.
Thus we find

\bea
V_u A&=&f_{r}{}^{s'}D_{s'}{}^{r}A+f_{r}{}^{s''}D_{s''}{}^{r}A
+f_{r''}{}^{s'}D_{s'}{}^{r''}A+  f_{r''}{}^{s''}\hat
D_{s''}{}^{r''}A \nn\\
&\phantom{=}&+ f_o D_o A + f_o'D'_o A + f_o'' D_o''A \eea

The first three terms involve the coset derivatives of the internal
flag manifold and the absence of their conjugates reflects the
H-analyticity of the field $A$. The fourth term takes care of the
transformation properties of $A$ under $SU(r)$ ($\hat
D_{r''}{}^{r''}=0$). The next two terms are left over from
$f_{r}{}^{s}D_{s}{}^{r}A$ and $ +f_{r'}{}^{s'}D_{s'}{}^{r'}A$ which
can only be non-trivial for the trace parts, and the final term
arises from removing the trace from $D_{r''}{}^{s''}$.  In the
above we have used the definition

\bea D_o={1\over p}D_r{}^r; \qquad f_o=f_r{}^r \eea

with similar definitions for primed and double-primed indices. The fact that we are working with $SU(N)$ then implies

\bea pD_o + qD'_o + rD''_o&=&0 \nn \\ f_o+f'_o + f''_o&=&0 \eea

A short calculation shows that

\bea D_{\a r}(\cV A)&=& {i\over2} \del_{\a\adt}\bar f^{\adt}_r D_o A\nn
\\ \bar D_{\adt}^{r'}(\cV A)&=& {i\over2}\del_{\a\adt} f^{\a r'} D'_o A
\eea

Since

\bea D_{\a i}K&=&-{i\over2} \del_{\a\adt}\bar f^{\adt}_i\nn \\ \bar
D_{\adt}{}^i\bar K&=&{i\over2}\del_{\a\adt} f^{\a i} \eea

we see that the desired transformation rule for $A$ is

\be
\d A=\cV A + (cK-c'\bar K)A \ee

where

\be
D_oA=cA;\qquad D_o'A=c'A \ee

This transformation rule can be rewritten in the form

\bea
\d A&=&f_{r}{}^{s'}D_{s'}{}^{r}A+f_{r}{}^{s''}D_{s''}{}^{r}A
+f_{r''}{}^{s'}D_{s'}{}^{r''}A+  f_{r''}{}^{s''}\hat
D_{s''}{}^{r''}A\nn\\
&\phantom{=}& + c(K + f_o-{p\over r}f_o'')A -c'(\bar
K-f_o'+{q\over r} f_o'')A \eea

The parameters $f_o-{p\over r}f_o''$ and $f_o'-{q\over r} f_o''$
are independent of the dilation and R-symmetry parameters, so that the
weights for these two transformations are determined by the parameters $c$ and
$c'$. Using the fact that

\bea D_o u_r&=&{N-p\over pN}u_r \nn\\ D_o u_{r'}&=& -{1\over
N}u_{r'}\nn\\ D_o u_{r''}&=& -{1\over N} u_{r''} \eea

together with similar formulae for $D_o'$ and $D_o''$, it is straightforward to
show that

\be
c=m_1-{m\over N},\qquad c'=-{m\over N} \ee

We therefore find

\be
L=m_1,\qquad R={2m\over N}-m_1 \ee

which is the correct series C result.

It is straightforward to adapt the above calculation to the special
cases $p=0$, $q=0$. For $N=4$ we have two possibilities, according to whether the superconformal group is chosen to be $PSU(2,2|4)$ or $SU(2,2|4)$. In both cases R-symmetry does not act on the coordinates, and in the former case it is absent. The parameter $D\cdot f+\bar D\cdot\bar f$ does act on the coordinates and is therefore not the R-parameter; it is the parameter of $U(1)_Y$ transformations in the terminology of \cite{i}. Since the corresponding generator does not belong to the superconformal algebra in either case the parameter $D\cdot f+\bar D\cdot\bar f$ must be set to zero.

For $PSU(2,2|4)$ the results derived above can be straightforwardly taken over provided one sets $R=0$. In the $SU(2,2|4)$ case, things are a little more subtle. It is possible to infer the existence of a real superfield parameter $\hat R$, say, whose leading component is the $R$ parameter and which satisfies

\be
D_{\a i} \hat R={\frac 14}\del_{\a\adt}\bar f^{\adt}_i
\ee

One can then define the chiral superfield

\be
K={\frac 14}\del\cdot F-i\hat R
\ee

and proceed as before. The only difference is that the R-transformation so defined does not act on the coordinates, although it does act on the superfield itself multiplicatively.

{\bf Acknowledgement}

This research was supported in part by PPARC SPG grant 613.

 \end{document}